\documentstyle[12pt,epsf]{article}   % Simple LaTeX, use with DINA4 as follows
\newlength{\dinwidth}
\newlength{\dinmargin}
\def\lsim{\mathrel{\rlap{\lower4pt\hbox{\hskip1pt$\sim$}}
    \raise1pt\hbox{$<$}}}                % less than or approx. symbol
\def\gsim{\mathrel{\rlap{\lower4pt\hbox{\hskip1pt$\sim$}}
    \raise1pt\hbox{$>$}}}                % greater than or approx. symbol
\begin{document}

\newpage

\vspace*{1cm}
\begin{center}  \begin{Large} \begin{bf}
On a Possibility to Determine \\
the Sign of the Polarized Gluon Distribution \\
\end{bf}
\end{Large}
  \vspace*{2cm}
  \begin{large}

W.-D. Nowak$^a$, A. V. Sidorov$^b$, M.V. Tokarev$^{c}$\\
  \end{large}
%\end{center}

\vskip 1cm

{\it $^a$ DESY-IfH Zeuthen, 15735 Zeuthen, Germany \\

\vskip 0.4cm
$^b$ Bogoliubov Laboratory of Theoretical Physics,
Joint Institute for Nuclear Research,\\
141980 Dubna, Moscow Region, Russia\\

\vskip 0.4cm
$^c$ Laboratory of High Energies,
Joint Institute for Nuclear Research,\\
141980 Dubna, Moscow Region, Russia\\}

\end{center}

\vskip 3.cm

\begin{abstract}
We investigate the possibility to draw conclusions on the sign of the
spin-dependent gluon distribution, $\Delta G(x, Q^2)$, from existing
polarized DIS data. The  spin-dependent parton distributions
 $\Delta u_v, \Delta d_v, \Delta {\bar u},
\Delta {\bar d}, \Delta {s}$, and $\Delta G$ are
constructed in the framework of
a phenomenological procedure  taking into account
some assumptions on signs of valence and  sea parton distributions
motivated
by 't Hooft's mechanism of quark-quark interaction induced by
instantons. The axial gluon anomaly and data on integral quark
contributions to the proton spin,
$\Delta \tilde u, \Delta \tilde d$, and $\Delta \tilde s$,
are also taken into account.
Predictions for the $x$- and  $Q^2$-dependencies
of the polarized proton and neutron structure functions,
$g_1^p$ and $g_1^n$, are compared to experimental data.
It is shown that the neutron structure function, $g_1^n$, is
especially  sensitive to the sign of $\Delta G(x, Q^2)$.
The results of our analysis supports the
conclusion  that this sign should be positive.
\end{abstract}

\newpage

{\section {Introduction}}

Parton distributions in the nucleon are of universal nature, hence
their parametrizations obtained from deep inelastic lepton-nucleon
scattering can be utilized for simulations of
processes outside lepton nucleon scattering; the polarized parton
distributions are especially useful to predict the behaviour of
$p p$ interactions with polarized proton beams to facilitate
future research programs at the RHIC, HERA and LHC colliders.

Recent results deep inelastic lepton-nucleon scattering experiments
at SLAC \cite{SLAC95,E143} and CERN \cite{SMC95}
on  spin-dependent structure functions
for proton and deuteron targets, $g_1^p$ and $g_1^d$,
stimulate the interest in determining the
spin-dependent gluon and quark distributions in a
polarized nucleon. Since a complete solution of this problem
is beyond the scope of
perturbative QCD and there are still no sufficiently precise
non-perturbative calculations available, the usual procedure is to
fit numerous parametrizations of both spin-independent and
spin-dependent parton distributions to the data. Up to now there
is no unique solution; the results depend in one or the other way
on the method used.

Polarized parton distributions can be extracted in an indirect manner
from doubly polarized deep inelastic lepton-proton
and  lepton-deuteron  scattering; the measurable observables
are the asymmetries  $A^p$ and $A^d$. The structure  function $g_1^p$
can be extracted from $A^p$ according to
\begin{equation}
g_1^p(x,Q^2)=A^p(x,Q^2)\cdot {F_2^p(x,Q^2)\over {2x(1+R(x,Q^2)}},
\end{equation}
where additional information on the unpolarized structure function
$F_2^p$ \cite{NMC} and on the ratio of longitudinal to transverse  photon
cross section $R(x,Q^2)={\sigma}_L/{\sigma}_T$ \cite{SLAC90} are required.
The deuteron structure function is determined in
a similar way taking into account appropriate nuclear corrections. \\
Since there is no practical way at present
to directly extract polarized parton distributions from experimental data
it is  important  to develop  flexible  procedures  to construct
these distributions incorporating relevant features of the data
as well as reasonable constraints derived from our present
theoretical understanding of the nucleon.

At present there is no strong argument favouring a positive or
negative sign of
the spin-dependent gluon distribution, $\Delta G(x, Q^2)$.
Several sets of spin-dependent parton distributions were constructed
utilizing rather different
approaches \cite{Brodsky,Gluck,Forte,TOK96} mostly assuming a positive
sign of $\Delta G$. Different parameter choices leading to a different
behaviour of $\Delta G$  at $x\rightarrow 1$ ($ G\uparrow \sim G\downarrow$,
\ $G \uparrow \gg G \downarrow$,\  $G \uparrow \ll G \downarrow$)
were studied in \cite{Stirling}. Both positive and negative values of
the sign  of  $\Delta G$ over a wide kinematical range $10^{-3}<x<1$
were considered in \cite{TOK96}.
A detailed NLO QCD analysis of the proton and deuteron data on
$g_1$ was performed in \cite{Forte} concluding that
the size of the gluon distribution drives the
perturbative evolution and, due to the fact that the SMC
and E143 data were taken at different values of $Q^2$,
the  polarized gluon  distribution turned out to be large and
positive.

The aim of
the present paper is to separate experimental observables being
sufficiently sensitive to allow a determination of the
sign of $\Delta G$. As we shall show later,
the  neutron structure function  $g_1^n$
seems to be one of those observables.  \\
To construct the spin-dependent parton distributions
a phenomenological method proposed in \cite{TOK96} is used.
This method incorporates some constraints on shape and sign of
parton distributions, it utilizes results on the quark contributions
to the nucleon spin obtained in other analyses, and the
effect of the axial anomaly is taken into account.
We study the $x$ and $Q^2$ dependence of $g_1^n(x,Q^2)$ for
two different scenarios: $\Delta G>0$ and $\Delta G<0$.
The calculated predictions are compared to experimental data; a
$\chi^2$ criterion is chosen to judge in which of the two scenarios
theoretical  curves are better describing the
experimental data on  $g_1^n(x,Q^2)$. Eventually, the choice for
a positive sign of $\Delta G$ will turn out to be the more likely
one, i.e. the polarized structure function of the neutron will be
shown to be sensitive to the sign of $\Delta G$.  \\

{\section{Method}}

The  spin-dependent proton structure function $g_1^p$ is
expressed through spin-dependent  parton distributions in a simple way
\begin{equation}
g_1^p(x,Q^2)={1\over 2}\cdot \{
{4\over 9}\Delta {\tilde u(x, Q^2)}+
{1\over 9}\Delta {\tilde d(x, Q^2)}+
{1\over 9}\Delta {\tilde s(x, Q^2)}\},
\end{equation}
where $\Delta q_f=q_f^{+}-q_f^{-}$, and the $q_f^{\pm}$ are the
probability distributions to find a quark having positive $(+)$ or
negative $(-)$ helicity relatively to positive proton helicity.
The neutron structure function $g_1^n(x,Q^2)$ can be written in a
similar form using the replacement
$\Delta {\tilde u} \leftrightarrow \Delta {\tilde d}$.
The valence distributions $\Delta {q_v}, \Delta {q_v}$ are then
obtained from $ \Delta {q_v}=\Delta {q}- 2\Delta {\bar{q}}$.    \\
Since in this paper we shall use  the
spin-dependent parton distributions constructed in \cite{TOK96}
we briefly describe in the following sections the main features
of the applied method.  \\

{\subsection{Shape of Parton Distributions}}

For the general form of a spin-dependent parton  distribution
$\Delta q_f$ we use
\begin{equation}
{\Delta} q_f= sign{(q_f)}\cdot x^{\alpha_f}\cdot
(1-x)^{\beta_f}
\cdot q_f, \ \ q_f=u_v,d_v,\bar u, \bar d, s, G.
\end{equation}
Here  $q_f$ is the spin-independent parton distribution,
$\alpha_f, \beta_f$ are free parameters which are to
be found by comparison with experimental data.
From the restriction
\begin{equation}
|\Delta q_f| <  q_f
\end{equation}
follows that both probability distributions  $q_f^+$,  $q_f^-$ as well
as their sum $q_f=q_f^{+} +q_f^{-}$ need to be positive; moreover
$\beta_f$ should not be negative. To avoid the latter
constraint we introduce a renormalised parton distribution
$ q_f^R=(1-x)^{\beta_f}\cdot q_f$. This leads to the following general
form of a spin-dependent parton distribution
\begin{equation}
{\Delta} q_f= sign{(q_f)}\cdot x^{\alpha_f} \cdot q_f^R.
\end{equation}
We note that since all presently available procedures to construct
both spin-independent and spin-dependent distributions do
imply fitting procedures and have consequently no unique solution.
Hence we believe that at present it is recommended to incorporate
general restrictions on $\Delta q_f$ like the one above; this
makes it easier to develop flexible schemes to construct the
helicity parton distributions $q_f^+$ and $q_f^-$, separately. \\

{\subsection{Signs of Parton Distributions}}

Up to now there exists neither a running experiment to directly
measure the polarized gluon distribution nor does the variety of
indirect analyses give a unique result. Hence there exist no
strong arguments on the sign of $\Delta G$. Our approach will be
to allow for both signs of $\Delta G$ and compare the quality of
our model-dependent predictions to the experimental data.

We note that a direct access to $\Delta G$ will be possible in
future experiments. Utilizing polarized protons at
RHIC for the (approved) experiments STAR and PHENIX \cite{RHIC} and,
possibly, for the suggested internal polarized target experiment
{\it HERA--$\vec{N}$} \cite{desy96-095} at HERA, the measurement of
$\Delta G$ seems feasible in the range $0.1 \leq x_{gluon} \leq 0.35$.
Also new doubly polarized lepton-nucleon scattering experiments
proposed at CERN \cite{compass} and suggested at SLAC
\cite{breton, bosted} might contribute very valuable information on
$\Delta G$.

For valence quark distributions the situation is much better defined;
we take the sign of $\Delta u_v$ as positive and that of $\Delta d_v$
as negative, respectively. This choice is motivated by the fact that the
dominant configuration in the proton wave function is
$u(\uparrow) u(\uparrow)d(\downarrow)$, here the arrow denotes the
quark  spin direction. The same choice is made in most analyses
of experimental data on quark contributions to the proton spin
\cite{SLAC95,E143,SMC95}, \cite{NMC88}-\cite{SMC94}.

We assume for signs of $\Delta \bar u$ ($\Delta \bar d$) to be
positive (negative).
This is motivated by 't\ Hooft's mechanism \cite{Hooft}
for the spin configuration  $u(\uparrow) u(\uparrow)d(\downarrow)$
which determines the dynamics of quark helicity flips.
The incoming left helicity  quark $q_L=(1+\gamma_5)q/2$  scattered
from zero modes in the instanton field leads to an outgoing
right helicity  quark  $q_R=(1-\gamma_5)q/2$.
Effective Lagrangians are constructed in \cite{Shif};
in the particular case of $N_f=2$ flavours it can be written as

\begin{equation}
{L} = \int d\rho\cdot n(\rho) ({4\over 3}{\pi}^2{\rho}^3)^2
\lbrace \bar u_R u_L \bar d_R d_L
\lbrack 1+ {3\over 32}(1-{3\over 4} \sigma_{\mu \nu}^u \sigma_{\mu \nu}^d )
\lambda_u^a \lambda_d^a \rbrack + (R\leftrightarrow L) \rbrace.
\end{equation}
Here $\rho$ is the size of instanton, $n(\rho)$ is the instanton density,
$\sigma_{\mu \nu}=i/4\cdot (\gamma_\mu \gamma_\nu
 - \gamma_\nu \gamma_\mu)$, and $\lambda^a$  are matrixes for $SU_c(3)$ group.
Once the left helicity quark scatters off an instanton it becomes a right
helicity one and a $q_R\bar q_R$ pair is created; the helicity of the sea
quarks being opposite to that of the initial quark. In other words,
the spin flip of the valence quarks $u^+$  and $d^-$  determines the sign of
the corresponding sea quark distributions - negative for $\Delta \bar d$ and
positive for $\Delta \bar u$. A negative sign
of $\Delta s$ is in agreement with the arguments mentioned above
and is supported by the results of several analyses of
$g_1^p$ data \cite{SLAC95,E143,SMC95}, \cite{NMC88}-\cite{SMC94}.  \\

{\subsection{Inclusion of Axial Anomaly }}

It was shown in \cite{Efremov} that the flavour-singlet axial current
\begin{equation}
A_{\mu}^0=\sum_{f=u,d,s} \bar q_f\gamma_{\mu}\gamma_{5} q_f
\end{equation}
diverges at the quark level due to the one-loop triangle anomaly
\begin{equation}
{\partial}^{\mu}A_{\mu}=
 {\alpha_s\over {\pi}}\cdot N_f\cdot
tr\lbrace F_{\mu \nu} \tilde F^{\mu \nu} \rbrace ,
\end{equation}
where
 $\tilde F_{\mu \nu}= {\epsilon}_{\mu \nu \beta \gamma }F^{\beta \gamma}$,
$ F_{\mu \nu}= \partial_{\mu}A_{\nu}- \partial_{\nu}A_{\mu}
+ \lbrack A_{\mu} A_{\nu}\rbrack $, $A_{\mu}= A_{\mu}^a\cdot \lambda^a$,
$\alpha_s$ is the strong coupling constant, and $N_f$ is the number of
flavours. The anomaly induces a mixing between the gluon
and the flavour singlet axial  current  of quarks. For this reason,
the helicity carried by each flavour undergoes a renormalization

\begin{equation}
\Delta \tilde q_f(x,Q^2) = \Delta q_f(x,Q^2)
 - {\alpha_s(Q^2) \over {2\pi}}\cdot  \Delta G(x,Q^2).
\end{equation}
It was suggested in \cite{Efremov} that the axial anomaly might
play a key role and modify the naive quark model predictions; hence
parton distributions will presumably become much more sensitive to the
sign of the polarized gluon distribution. Consequently, the
spin-dependent structure functions $g_1^p$ and $g_1^n$ would become
more sensitive to $\Delta G$, as well.  \\

{\subsection{Integral Parton Contributions to the Proton Spin}}

A further input required to our analysis is the total contribution of
each quark species to the proton spin. We utilize the results of a
recent analysis \cite{Ellis} of the structure functions $g_1^p$  and
$g_1^d$ from SMC and SLAC data incorporating $3^{rd}$ order pQCD corrections
to the Bjorken sum rule. The relative quark contributions to the
proton spin were determined as
$\Delta \tilde u = 0.83 \pm 0.03, \Delta \tilde d = -0.43 \pm 0.03,
\Delta \tilde s=-0.10\pm 0.03$ at a renormalization scale
$Q^2_0 = 10$ (GeV/c)$^2$. Using these values and the definition
\begin{equation}
\int_0^1 \Delta {\tilde q_f}(x,Q_0^2) dx
= \Delta \tilde f,\ \ f= u, d, s
\end{equation}
the free parameters $\alpha_f, \beta_f$ in the parametrization of our
spin-dependent parton distributions
 $\Delta u_V, \Delta d_V, \Delta {\bar u},
\Delta {\bar d},\Delta {s},\Delta G$ were determined in \cite{TOK96}. \\

\vskip 0.5cm

{\section {Results and Discussion}}

In fig.~1~(a,b) and 2~(a,b) the $x$-dependence of $g_1^p$ and $g_1^n$
is shown for different parametrizations of parton distributions constructed
with $\Delta G >0$ (a) and   $\Delta G <0$ (b).
The dashed, solid and dotted lines  correspond
to the parameters  ${\alpha}_f, {\beta_f}$  taken from
 Table 1-3 and 4-6 of Ref.~\cite{TOK96}, respectively.

From fig.~1~(a,b) is seen that all theoretical curves for the {\it proton}
structure function $g_1^p$ are in reasonable agreement
with experimental data \cite{SLAC95,NMC88,SMC94}, i.e. there seems to
be no apparent sensitivity to the sign of $\Delta G$. In contrast,
from fig.~2~(a,b), displaying experimental data and theoretical curves
for the {\it neutron} structure function $xg_1^n$, one can deduce
a certain dependence of the theoretical curves on the sign of $\Delta G$
in the range $0.1~<~x~<~0.3$. Hence there is some hope that $xg_1^n$
exhibits a certain sensitivity to the sign of $\Delta G$.

Fig.~3~(a,b) shows the $x$-dependence of the {\it proton} structure function
$xg_1^p(x,Q^2)$ at different
values of four-momentum transfer, $Q^2 = 1, 10, 100$ (GeV/c)$^2$.
The $Q^2$ behaviour of $xg_1^p$ appears qualitatively different for
$\Delta G >0$ and $\Delta G < 0$, respectively.
In the first case the maximum of the  curve is moved to lower $x$
with increasing $Q^2$, in the second one the position of the maximum
is not affected. If $\Delta G>0$ the prediction {\it increases} with $Q^2$
for $x~<~0.01$. If $\Delta G<0$, the prediction {\it decreases} with $Q^2$
over the full $x$-range.

Fig.~4~(a,b) displays the $x$-dependence of the {\it  neutron} structure
function $xg_1^n(x,Q^2)$ in the same
fashion, i.e. for $Q^2 = 1, 10$ (GeV/c)$^2$. If $\Delta G>0$ the
differences for different $Q^2$ appear mainly at very low $x$-values
and, in addition, at moderate $x \simeq 0.3$. This sensitivity to the
sign of $\Delta G$ is to weak for present experimental errors, however,
it might be used later when more precise data on $g_1^n(x,Q^2)$ should
become available. For $\Delta G<0$ one observes a rather
strong $Q^2$-dependence at lower $x$-values and a somewhat
characteristic dip at higher $x$, its position being
almost independent on $Q^2$.

To be closer to the presently available $Q^2$-values we show in fig.~5
the $x$-dependence of $xg_1^n(x,Q^2)$ at $Q^2 = 1, 3, 5, 10$ (GeV/c)$^2$
together with the presently available experimental data.
(Due to the experimental errors  the different ordinate
is choosed  in fig.~5(d) than in fig.~5(a,b,c).)
The behaviour of $xg_1^n$ on $Q^2$
is qualitatively and quantitatively different
for the two scenarios $\Delta G >0$ and $\Delta G < 0$,
especially at low $Q^2$. Apparently, in the range  $x<0.1$ the
experimental data on $g_1^n$ at
$Q^2~<~10~$(GeV/c)$^2$ should be able to discriminate between positive
and negative sign of the polarized gluon distribution.

We apply a $\chi^2$ criterion to quantitatively distinguish between
the two scenarios by comparing our constructed parton distributions to the
experimental data sets from SLAC and CERN \cite{E143,SMC95,SLAC93}.
The obtained results  are summarized in Table 1.
There the references for experimental data on $g_1^n$,
the average $Q^2$ values, and the
number of experimental points are shown  in column~1, 2, and 3,
respectively. The 'all' in col.~2 takes into account that each
individual experimental
point was measured at another average $Q^2$, i.e. here the
$\chi^2$ is calculated using in the theoretical calculation
the correct average $Q^2$-value at each $x$-point.
The corresponding kinematically
accessible ranges are $1.1 \div 5.2$ (GeV/c)$^2$ for E142
and $1.3 \div~48.7$ (GeV/c)$^2$ for SMC. This method seems to us the
closest possible description of the data by a theoretical calculation,
hence we expect the $\chi^2$ values obtained for the 'all'
comparison to yield the best possible separation.

\vskip 0.5cm

\begin{center}

\begin{tabular}{||c||c|c|c|c||} \hline \hline
 Experiment          & $<Q^2>$& data & $\chi ^2$ / ndf & $\chi ^2$ / ndf   \\
                     & $(GeV/c)^2$&points&${\Delta G>0}$&${\Delta G<0}$
 \\ \hline \hline
  E142 \cite{SLAC93}  &   2  &   8   &   1.20     &   2.05       \\  \hline
  E143 \cite{E143}    &   3  &   9   &   0.89     &   1.41       \\  \hline
  SMC  \cite{SMC95}   &  10  &   12  &   1.28     &   1.63       \\  \hline
  HERMES \cite{HERM96}&   3  &   8   &   0.86     &   1.20       \\  \hline
  E142 \cite{SLAC93}  &  all &   8   &   1.45     &   2.30       \\  \hline
  SMC  \cite{SMC95}   &  all &   12  &   1.35     &   2.41       \\  \hline \hline
\end{tabular}

\end{center}

{\it {Table 1. $\chi^2$ comparison between theoretical predictions,
calculated for the two scenarios $\Delta G>0$ and $\Delta G<0$,
and experimental data on $g_1^n(x,Q^2)$.
}\\[0.5cm]}

From table 1 one can see that for every data set the
$\chi^2$ per degree of freedom is significantly
better in the case $\Delta G >0$
compared to the case $\Delta G < 0$.
These results can be considered as clear
quantitative evidence that the case
$\Delta G >0$  is the more likely scenario compared to the case
$\Delta G < 0$.  \\
We note that our result supports the conclusion on a
positive sign of $\Delta G >0$ obtained recently by a NLO QCD fit to
$g_1$ proton and deuteron data \cite{Forte}.

Finally we present in table 2 our results for the integral
quark -- $\Delta \Sigma $ -- and gluon -- $\Delta g$ -- contributions
to the proton spin calculated with the low limit $x_{min}=10^{-3}$.
Whereas in the more likely scenario $\Delta G >0$ the quark
contribution $\Delta \Sigma$ appears to be almost stable with $Q^2$
it drops by almost a factor of 2 when increasing $Q^2$ from 3 to 10
(GeV/c)$^2$ in the less likely case $\Delta G <0$. In both scenarios
$\Delta g$ rises by about 10\% within the same $Q^2$ range.

\begin{center}
\begin{tabular}{||c||c|c||c|c||} \hline \hline
\baselineskip =3.\baselineskip
${Q_0^2} $      &\multicolumn{2}{c||}{ $\Delta \Sigma $}
                &\multicolumn{2}{c||}{ $\Delta g      $}\\ \cline{2-5}
${(GeV/c)}^2$   &$\Delta G>0$ &$\Delta G<0$ &$\Delta G>0$ &$\Delta G<0$\\ \hline \hline
 3      & 0.290 & 0.520 &1.78 &-3.01  \\
 5      & 0.293 & 0.420 &1.86 &-3.20  \\
10      & 0.298 & 0.296 &1.95 &-3.41  \\ \hline \hline
\end{tabular}
\end{center}

{\it {Table 2. Integral quark --  $\Delta \Sigma$ -- and gluon  --
    $\Delta g$ -- contributions to the proton spin,
calculated from the constructed polarized parton distribution functions
for the two scenarios $\Delta G>0$ and $\Delta G<0$.
}\\[0.5cm]}

{\section {Conclusions}}

The possibility to draw conclusions on a positive or negative sign of
the polarized gluon distribution $\Delta G(x,~Q^2)$ was studied using a
phenomenological procedure to construct spin-dependent parton
distributions. The method includes some constraints
on the signs of valence and  sea quark distributions,
takes into account the axial gluon anomaly and utilizes results
on integral contributions to the nucleon spin,
$\Delta \tilde u, \Delta \tilde d, \Delta \tilde s$.
Investigating the $x$- and  $Q^2$-dependencies
of the structure functions $g_1^p$ and $g_1^n$ constructed by this method
we introduce a $\chi^2$ criterion to discriminate between the two scenarios
obtained for $\Delta G~>~0$ and $\Delta G~<~0$, respectively.
The neutron structure function turned out to be sufficiently sensitive
to the sign of $\Delta G(x, Q^2)$, even at the present level of only moderate
experimental errors. The results of our analysis strongly support the
conclusion  that the sign of $\Delta G(x, Q^2)$ is positive.
New data on the neutron structure function $g_1^n$
from the latest SLAC experiments and from HERMES at DESY will undoubtedly
allow to draw a more definite conclusion on the sign of  the polarized
gluon distribution.  \\

{\section*{Acknowledgements}}

This work was partially  supported
by  Grants of the Heisenberg-Landau Program for 1996
and of the Russian  Foundation
of Fundamental Research  under  No. 95-02-05061 and No. 96-02-18897.  \\

%\newpage

\newpage

\begin{center}
\vskip -1.cm
%\parbox{8cm}{\epsfxsize=8.cm \epsfysize=8.cm \epsfbox[5 5 500 500]
\parbox{8cm}{\epsfxsize=8.cm \epsfysize=8.cm \epsfbox[35 5 535 500]
{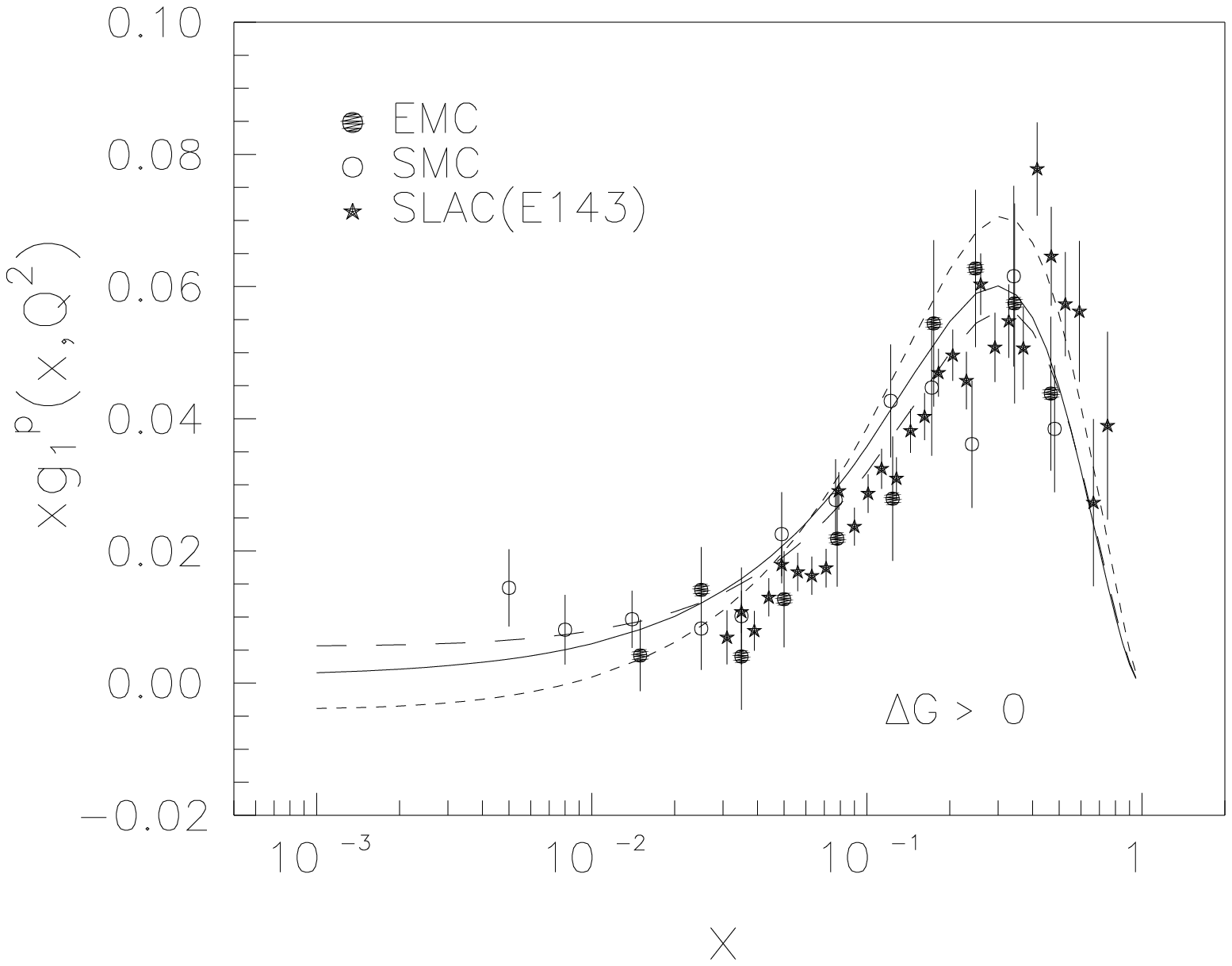}{}}

\hspace*{0.cm} a)

\parbox{8cm}{\epsfxsize=8.cm \epsfysize=8.cm \epsfbox[35 5 535 500]
%\parbox{8cm}{\epsfxsize=8.cm \epsfysize=8.cm \epsfbox[5 5 500 500]
{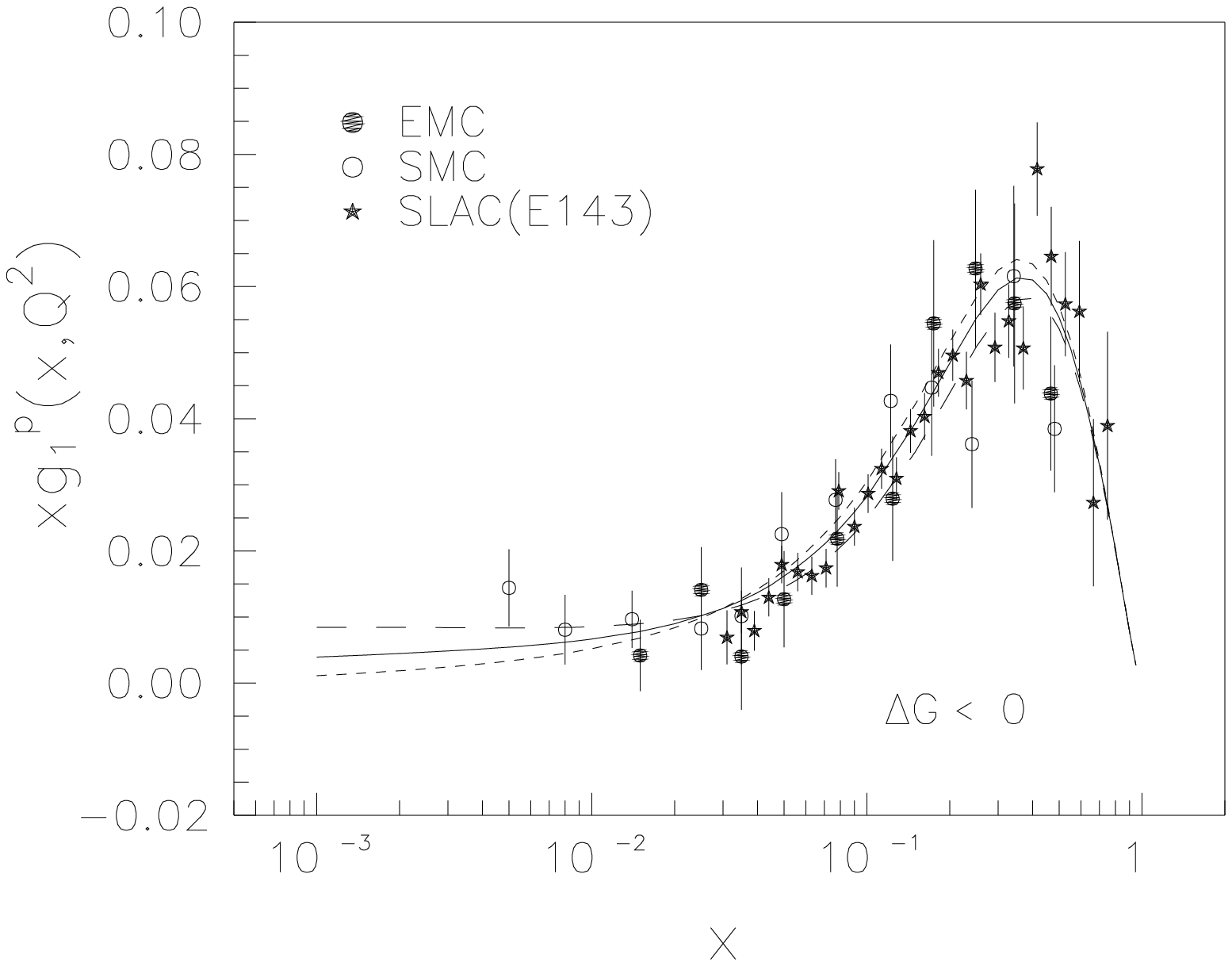}{}}

\vskip 0.5cm

\hspace*{0.cm} b)

\end{center}

{\bf Figure 1.}
 Deep-inelastic  proton structure function  $xg_1^p (x,Q^2)$.
Experimental data: $\star $ - \cite{SLAC95},
 $\bullet$ - \cite{NMC88},
$\circ$ - \cite{SMC94}.
Theoretical curves: (a) - $\Delta G > 0 $ and (b) - $\Delta G <0 $
at $Q^2=10\ (GeV/c)^2$.
Parametrizations of parton distributions:
 -- -- -- ,
 -------- ,  --- ---  are  taken from
 Tables 1-3 and
 Tables 4-6   \cite{TOK96}, respectively.

\begin{center}
\vskip -2cm
%\parbox{8cm}{\epsfxsize=8.cm \epsfysize=8.cm \epsfbox[5 5 500 500]
\parbox{8cm}{\epsfxsize=8.cm \epsfysize=8.cm \epsfbox[35 5 535 500]
{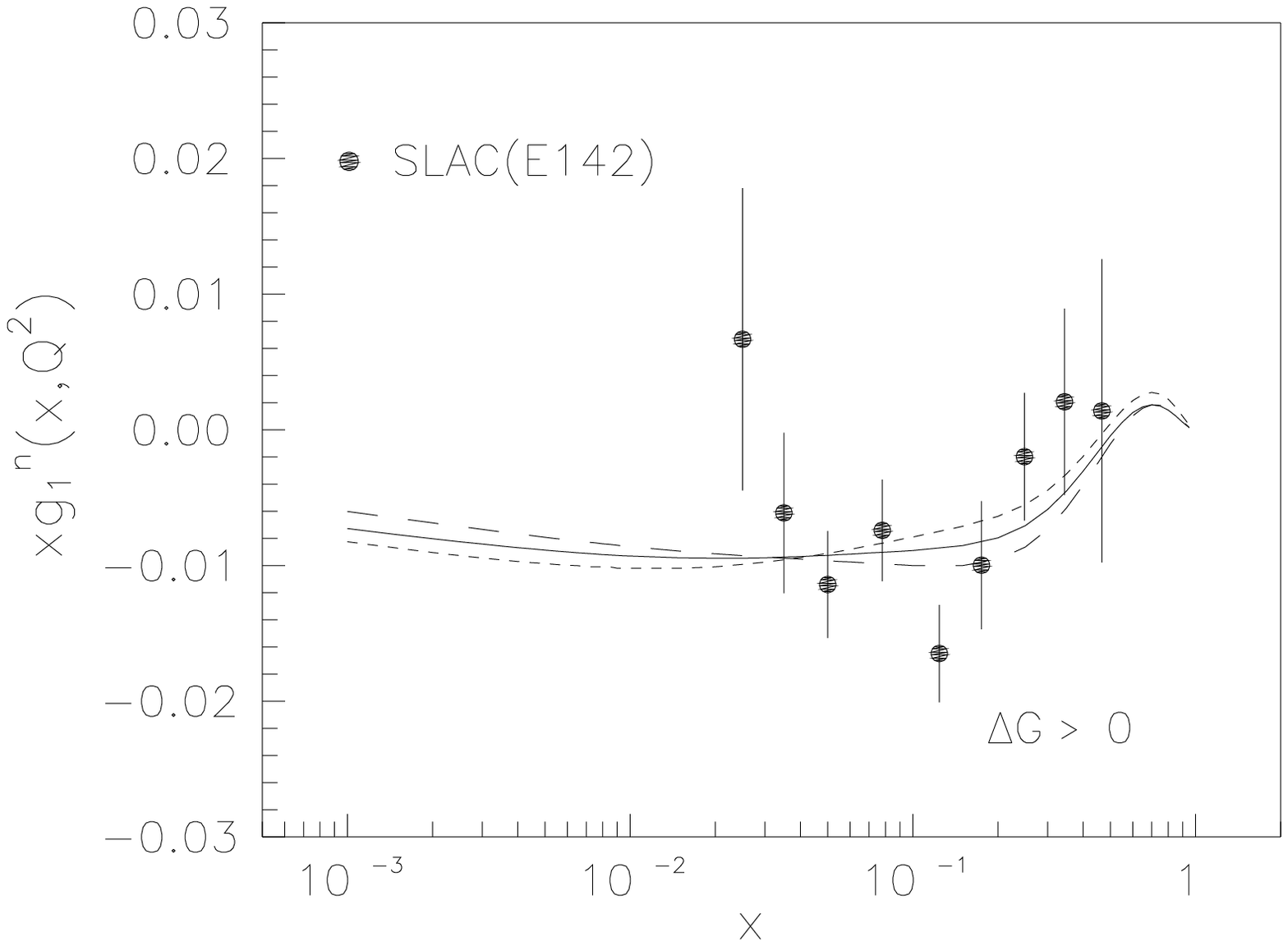}{}}

\hspace*{0.cm} a)

\parbox{8cm}{\epsfxsize=8.cm \epsfysize=8.cm \epsfbox[35 5 535 500]
{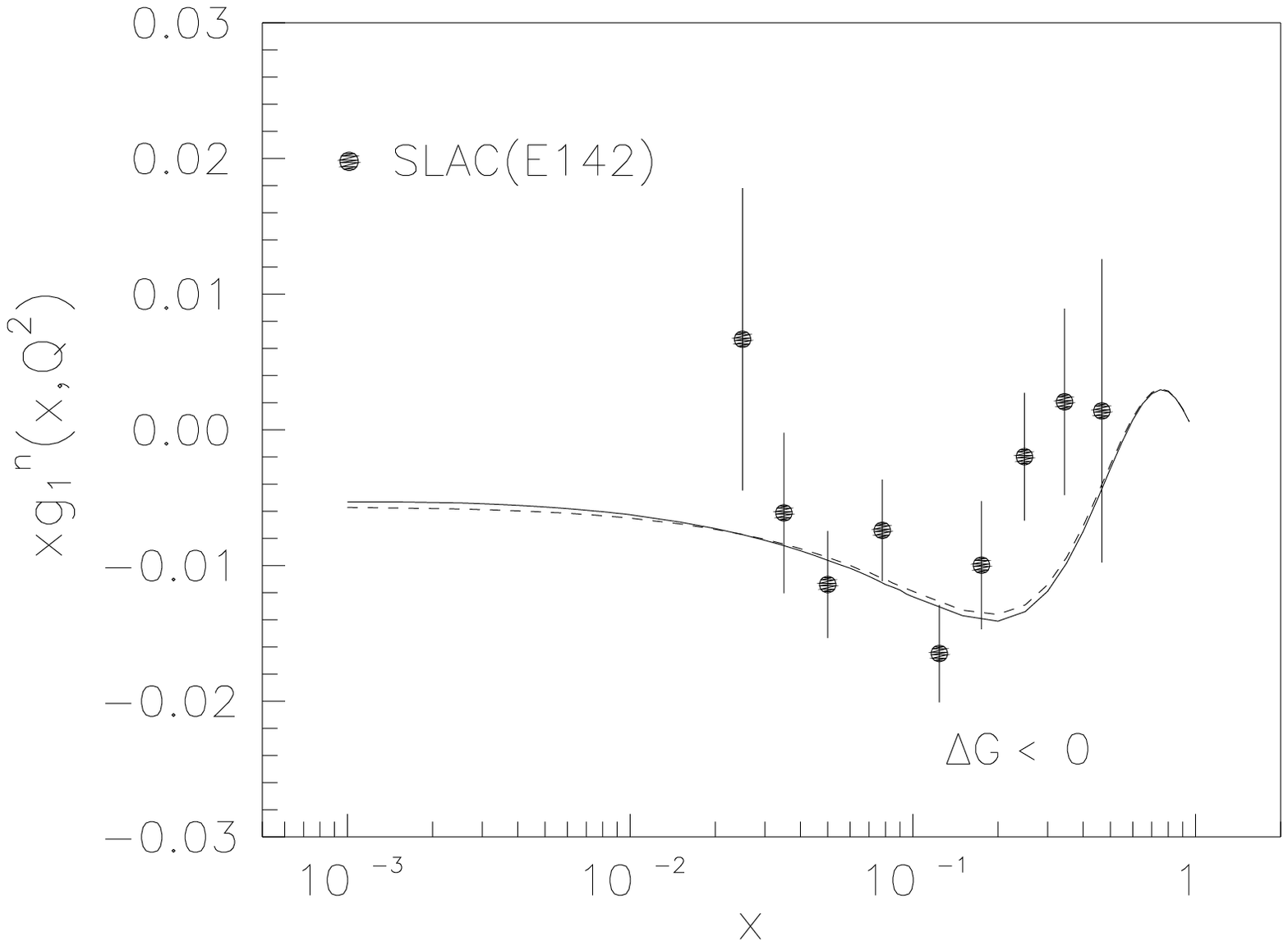}{}}

\vskip 0.5cm

\hspace*{0cm} b)

\end{center}

{\bf Figure 2.}
 Deep-inelastic  neutron  structure function  $xg_1^n (x,Q^2)$. Experimental
 data:  $\bullet $ - \cite{SLAC93}.
Theoretical curves: (a) - $\Delta G > 0 $ and (b) - $\Delta G <0 $
at $Q^2=10\ (GeV/c)^2$.
Parametrizations of parton distributions: -- -- -- ,
 -------- ,  --- ---  are  taken from
 Tables 1-3 and   Tables 4-6   \cite{TOK96}, respectively.

\newpage
\begin{center}
\vskip -2cm
\parbox{8cm}{\epsfxsize=8.cm \epsfysize=8.cm \epsfbox[35 5 535 500]
{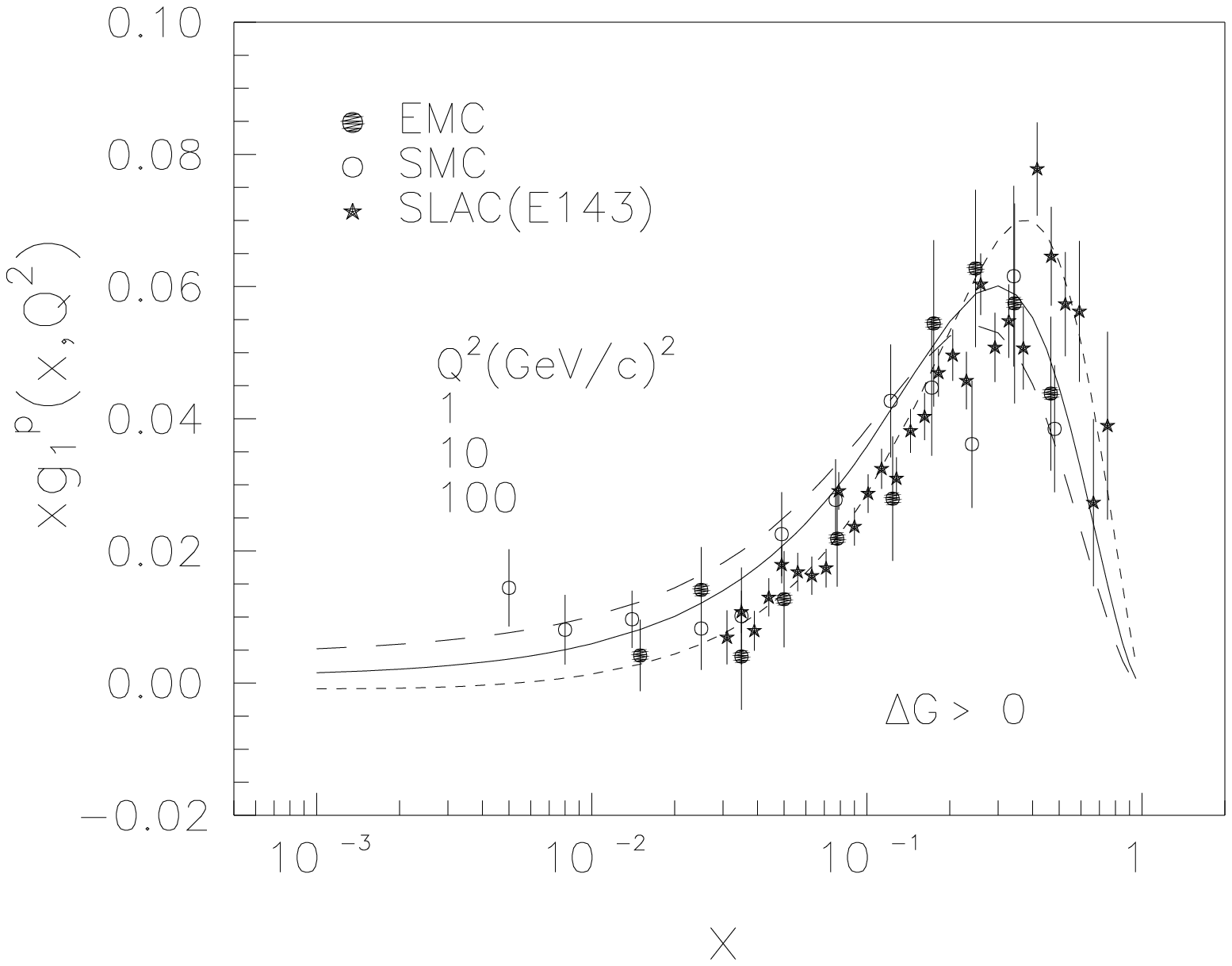}{}}

\hspace*{0.cm} a)

\parbox{8cm}{\epsfxsize=8.cm \epsfysize=8.cm \epsfbox[35 5 535 500]
{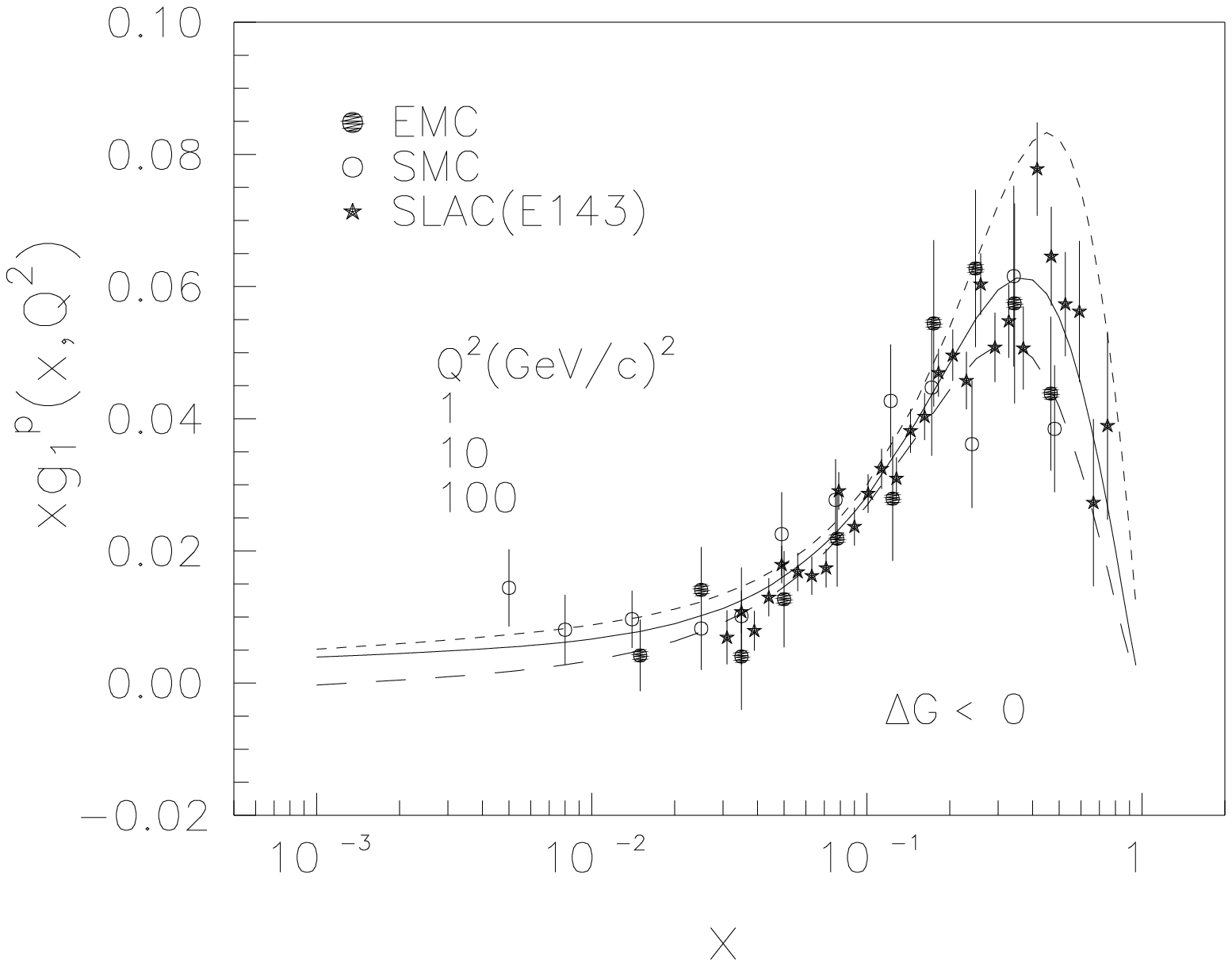}{}}

\vskip 0.5cm

\hspace*{0.cm} b)
\end{center}

{\bf Figure 3.}
 Deep-inelastic  proton structure function  $xg_1^p (x,Q^2)$.
Experimental data:   $\star $ - \cite{SLAC95},
$\bullet$ - \cite{NMC88},
$\circ$ - \cite{SMC94}.
Theoretical curves: (a) - $\Delta G > 0 $, \ (b) - $\Delta G <0 $
and \ -- -- --  - 1\  $(GeV/c)^2$,
\ --------  - 10\ $(GeV/c)^2$, \ --- ---  - 100\ $(GeV/c)^2$.
Parametrizations of parton distributions
 are  taken from Tables 2  and 5 \cite{TOK96}.

\begin{center}
\vskip -2cm
\parbox{8cm}{\epsfxsize=8.cm \epsfysize=8.cm \epsfbox[35 5 535 500]
{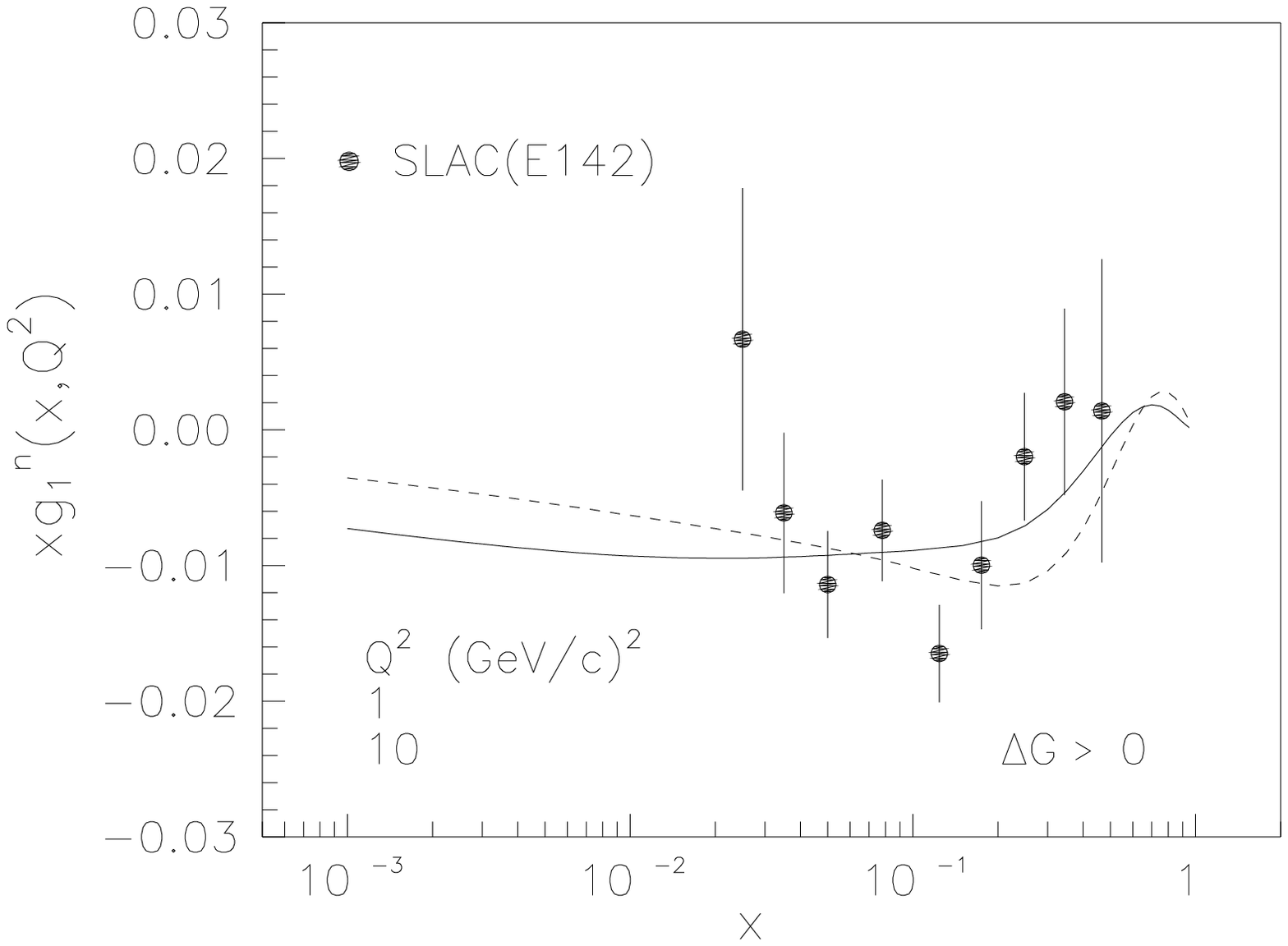}{}}

\hspace*{0.cm} a)

\parbox{8cm}{\epsfxsize=8.cm \epsfysize=8.cm \epsfbox[35 5 535 500]
{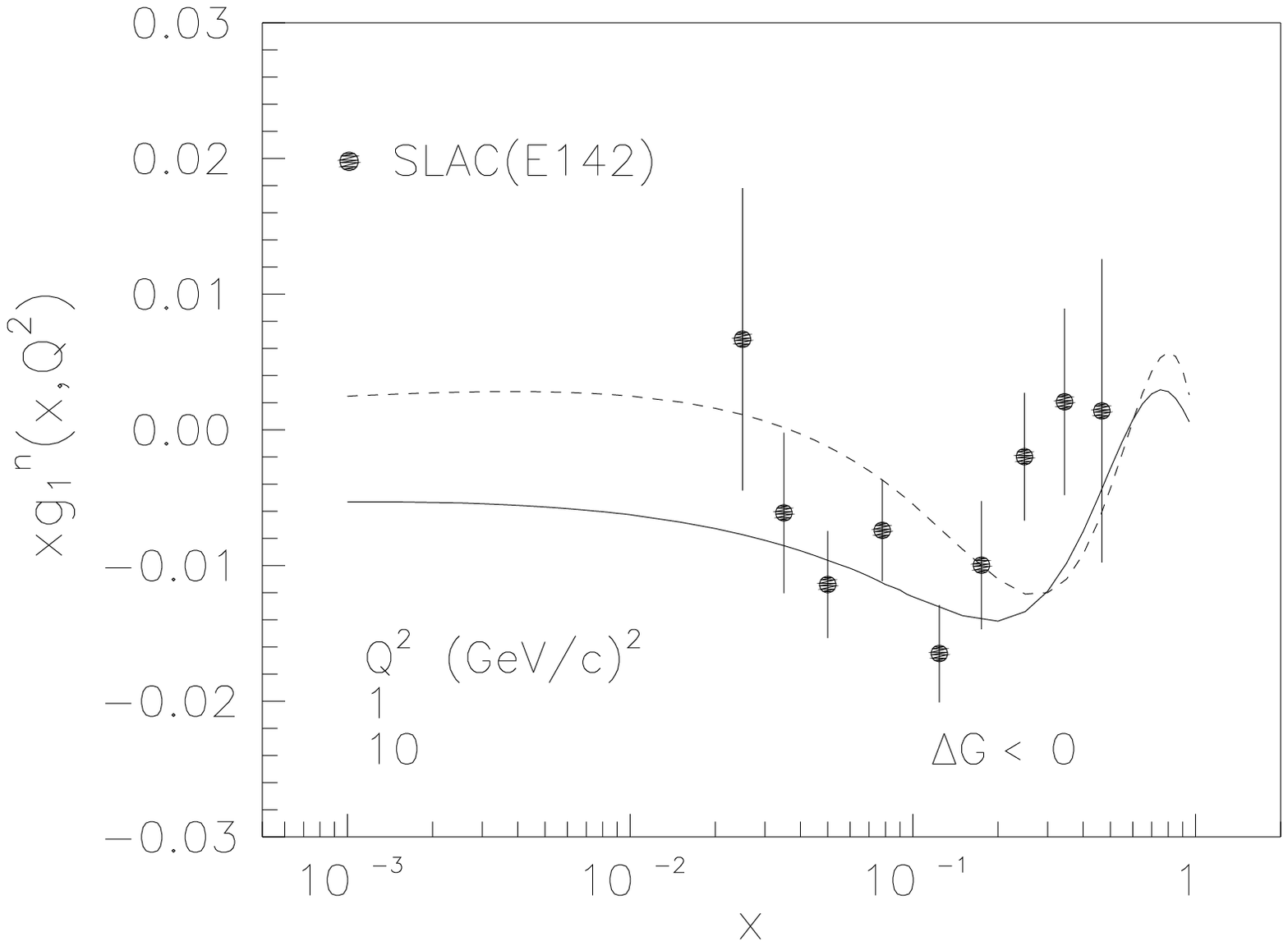}{}}

\vskip 0.5cm

\hspace*{0.cm} b)
\end{center}

{\bf Figure 4.}
 Deep-inelastic  neutron structure function $xg_1^{n}$.
Experimental data: $\bullet $ - \cite{SLAC93}.
 Theoretical curves: (a) - $\Delta G > 0 $, \  (b) - $\Delta G <0 $
and \ -- -- --  - 1\  $(GeV/c)^2$, \ --------  - 10\ $(GeV/c)^2$.
Parametrizations of parton distributions
 are  taken from Tables 2  and 5 \cite{TOK96}.

%*************************************************************************
\newpage
\begin{center}
\vskip -2cm
\parbox{8cm}{\epsfxsize=8.cm \epsfysize=8.cm \epsfbox[35 5 535 500]
{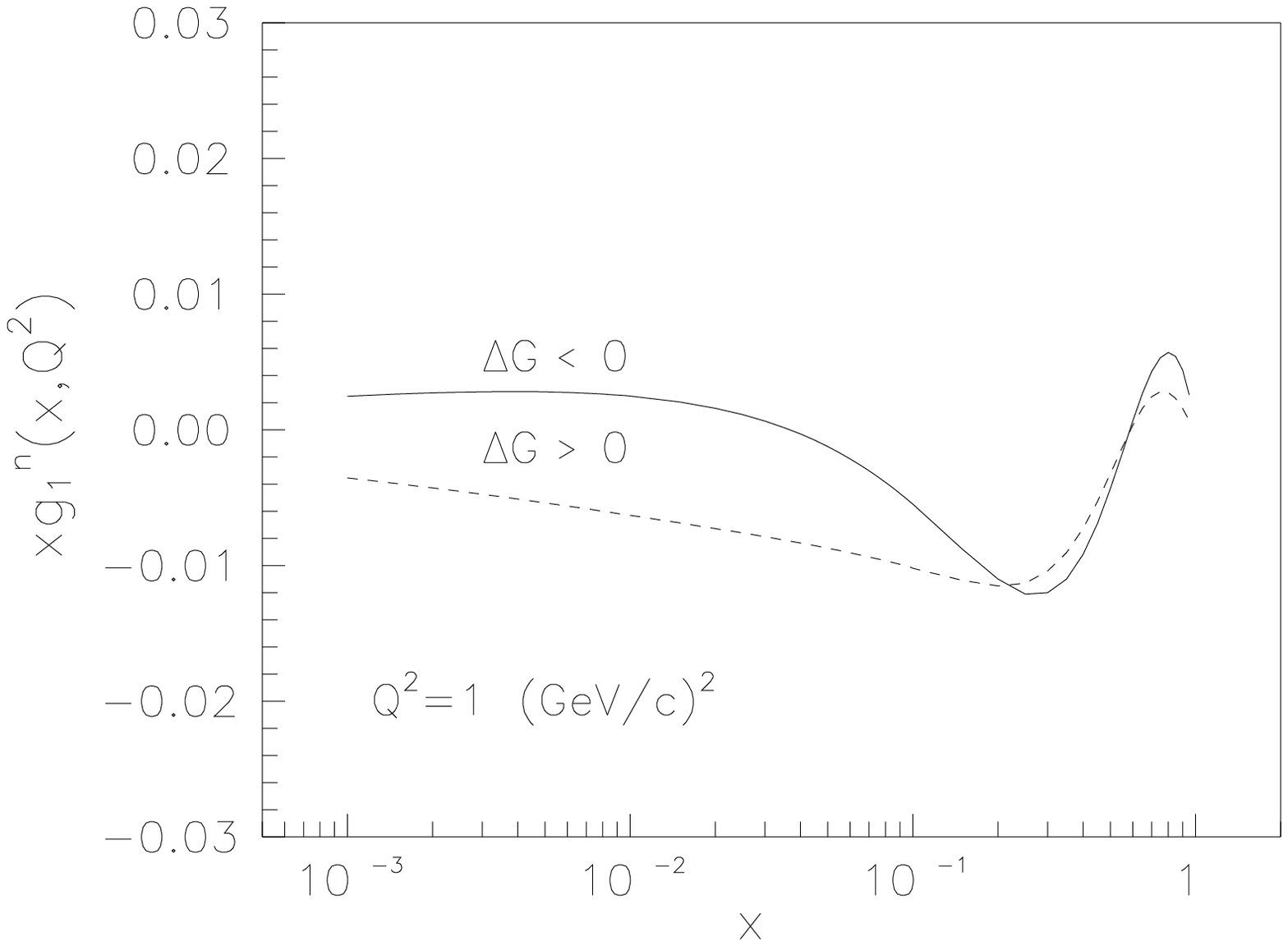}{}}

\hspace*{0. cm} a)

\parbox{8cm}{\epsfxsize=8.cm \epsfysize=8.cm \epsfbox[35 5 535 500]
{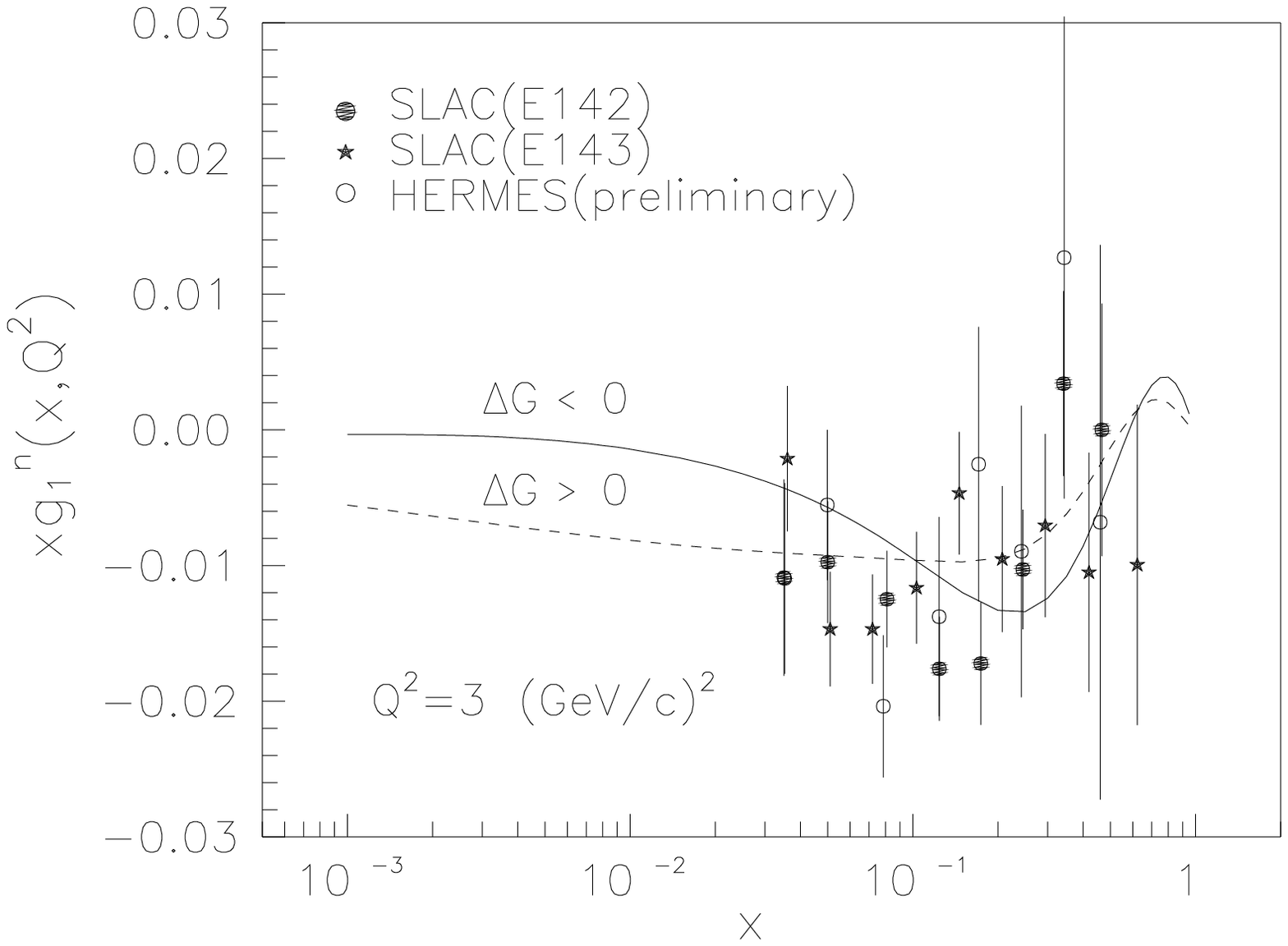}{}}

\vskip 0.5cm
\hspace*{0. cm} b)
\end{center}

{\bf Figure 5.}
 Deep-inelastic  neutron structure function  $xg_1^n (x,Q^2)$.
Experimental data:    $\star$ - \cite{E143},
$\triangle$ - \cite{SMC95},
 $\bullet$ - \cite{SLAC93},
$\circ$ - \cite{HERM96}.
Theoretical curves:  -- -- --   - $\Delta G > 0 $,
\ -------   - $\Delta G <0 $ at $Q^2 = 1,3,5,10\ (GeV/c)^2$.
Parametrizations of parton distributions
 are  taken from Tables 2  and 5 \cite{TOK96}.

\newpage

\begin{center}
\vskip -2cm
\parbox{8cm}{\epsfxsize=8.cm \epsfysize=8.cm \epsfbox[35 5 535 500]
{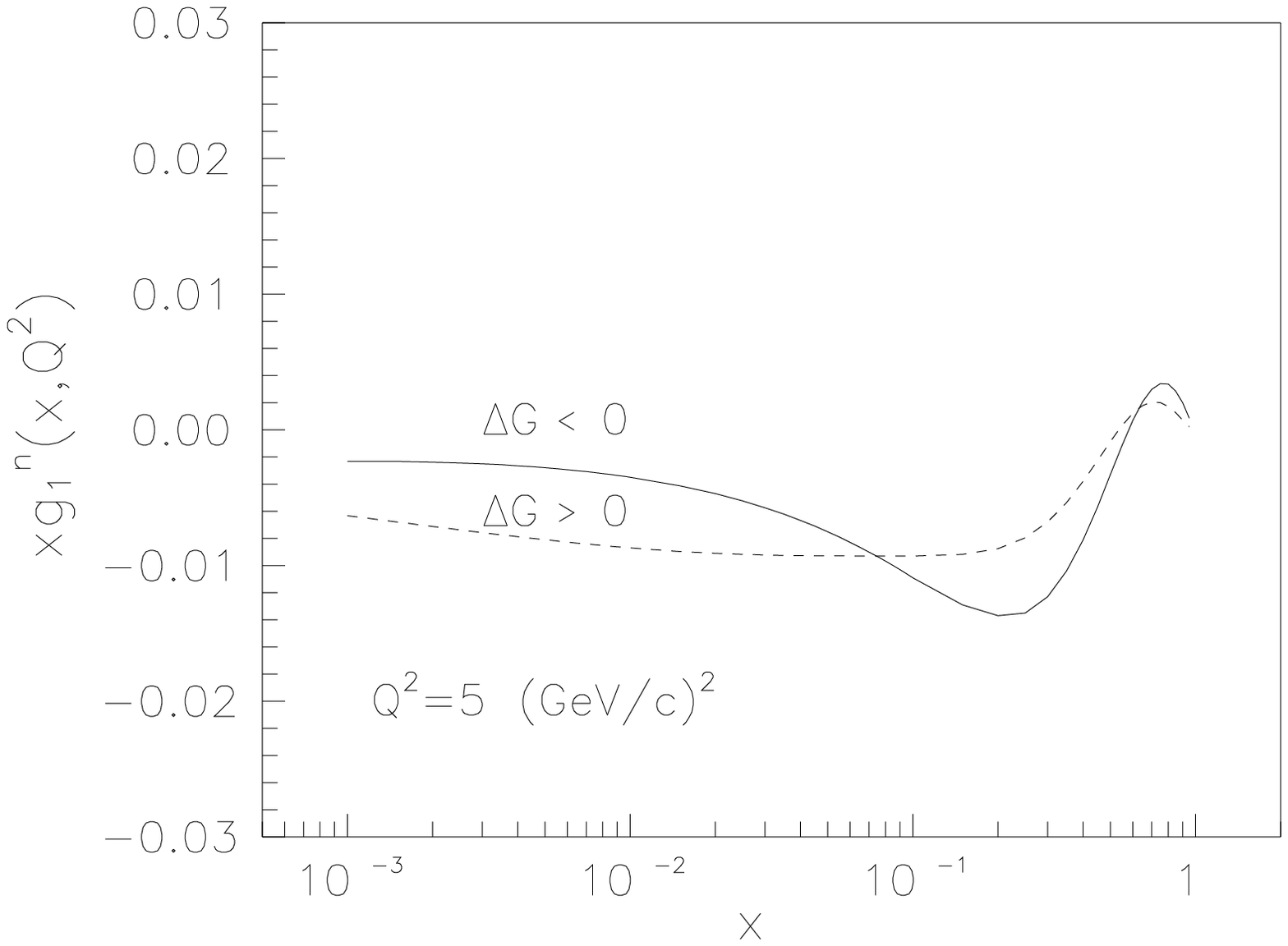}{}}
%{w11e154.ps}{}}

\hspace*{0.cm} c)

\parbox{8cm}{\epsfxsize=8.cm \epsfysize=8.cm \epsfbox[35 5 535 500]
%{w121.ps}{}}
{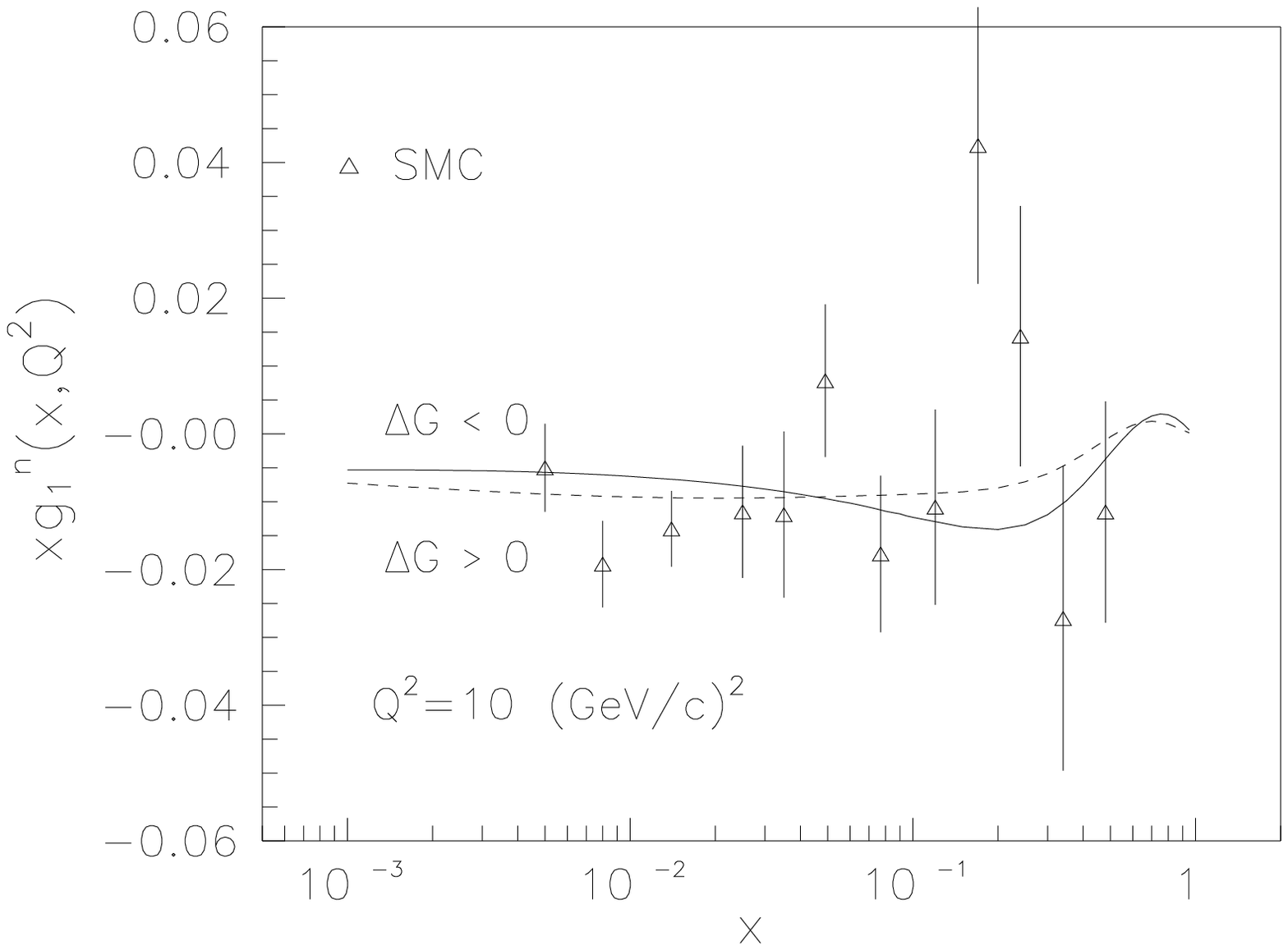}{}}

\vskip 0.5cm

\hspace*{0.cm} d)

\vskip 0.5cm
{\bf Figure 5.}   Continued

\end{center}

\end{document}